\documentclass[prd,12pt,tightenlines,nofootinbib]{revtex4}
\usepackage{amssymb,amsthm}
\usepackage{amsmath,amssymb}
\newcommand{\be}{\begin{equation}}
\newcommand{\ee}{\end{equation}}
\newcommand{\bea}{\begin{eqnarray}}
\newcommand{\eea}{\end{eqnarray}}
\newcommand{\del}{\partial}
\newcommand{\abs}[1]{\left| #1 \right|} % for absolute value

\usepackage{expl3}
\ExplSyntaxOn
\newcommand\latinabbrev[1]{
  \peek_meaning:NTF . {% Same as \@ifnextchar
    #1\@}%
  { \peek_catcode:NTF a {% Check whether next char has same catcode as \'a, i.e., is a letter
      #1.\@ }%
    {#1.\@}}}
\ExplSyntaxOff

\def\ie{\latinabbrev{i.e}}    

\begin{document}

\title{On Determining the Running Coupling from the Effective Action}

\author{D.G.C. McKeon$^{b,c}$}\email{dgmckeo2@uwo.ca}
\author{A. Patrushev$^{b}$}\email{apatrush@uwo.ca}
\affiliation{$^b$Department of Applied Mathematics, University of Western Ontario, London Canada N6A 5B7}
\affiliation{$^c$Department of Mathematics and Computer Science, Algoma University, Sault St.Marie Canada P6A 2G4, Canada}

\begin{abstract}

The conformal anomaly has provided an expression for the effective action of gauge theories in the presence of a strong background field in terms of the running coupling constant.  We exploit this result to find a novel expansion for the running coupling.
\end{abstract}

\maketitle

\section{Introduction}
It has been long known that the introduction of a renormalization scale $\mu$ leads to a conformal anomaly.
More explicitly, the trace of the energy-momentum tensor is no longer zero but rather is proportional to the renormalization group $\beta$-function \cite{Anom}. 
From this result, one can show that the effective action for a gauge theory can be written in terms of the running gauge coupling when considered as a function of a strong background field \cite{b2}. 
At the same time, the effective action satisfies the renormalization group equation, which leads to explicit summation of all its leading-log (LL), next-to-leading-log (NLL) etc. contributions \cite{b3}.
In this paper we exploit these two different expressions for the effective action to obtain a novel expression for the running gauge coupling. We relate this new expansion to one previously derived 
by systematically solving the usual differential equation for the running coupling.

\section{The Running Coupling and the Effective Action}

If the effective Lagrangian $L$ is treated as a function of $\mu$ (the renormalization scale), $F_{\mu\nu}$ (the constant background field strength) and $\lambda$ (the gauge coupling),
then we have the renormalization group equation:
\begin{equation}\label{eq1}
\mu\frac{dL}{d\mu}=\left(\mu\frac{\del}{\del\mu}+\beta(\lambda)\frac{\del}{\del\lambda}+\gamma(\lambda) F_{\mu\nu}\frac{\del}{\del F_{\mu\nu}}\right)L(\lambda,F_{\mu,\nu},\mu)=0.
\end{equation}
Since $\lambda F_{\mu\nu}$ is not renormalized [4] it follows that $\beta(\lambda)=-\lambda\gamma(\lambda)$ and equation \eqref{eq1} becomes
\begin{equation}\label{eq2}
\left[\mu\frac{\del}{\del\mu}+\beta(\lambda)\left(\frac{\del}{\del\lambda}-\frac{2}{\lambda}\Phi\frac{\del}{\del\Phi}\right)\right]L=0,
\end{equation}
where $\Phi=F_{\mu\nu}F^{\mu\nu}$.

For strong background fields (  \ie $\lambda \Phi \gg \mu^2$)
\begin{equation}\label{eq3}
L=\sum_{n=0}^{\infty}\sum_{m=0}^{\infty} T_{n,m}\lambda^{2n}t^m\Phi = 
\sum_{n=0}^\infty \,S_n(\lambda^2t)\lambda^{2n}\Phi
\end{equation}
where $t=\frac{1}{4}\ln\left(\frac{\lambda^2\Phi}{\mu^4}\right)$ [5] and  $S_n(\lambda^2 t)=\sum_{m=0}^{\infty}T_{n+m,m}(\lambda^2t)^m$ ($n=0$ is LL, $n=1$ is NLL etc.). Eq. \eqref{eq2} leads 
to the nested equations ($n=0,1,2 \ldots$)

\begin{equation}\label{eq4}
-\frac{d}{d\xi}S_n(\xi)+2\sum_{\rho=0}^nb_{2\rho+3}\left[\xi \frac{d}{d\xi}+(n-\rho-1)\right] S_{n-\rho}=0
\end{equation} 
where $\beta(\lambda)=\sum_{\rho=0}^{\infty} b_{2\rho+3}\lambda^{2\rho+3}$ and $\xi=\lambda^2t$. The boundary condition for these equations is $S_n(\xi=0)=T_{n,0}$. Solutions for $n=0,1,2$ 
are respectively given by
\begin{subequations}\label{eq5}
\begin{equation}\label{eq5a}
S_0=-T_{0,0} w 
\end{equation}
\begin{equation}
S_1=\frac{T_{0,0}b_5}{b_3}\ln\abs{w}+T_{1,0}
\end{equation}
\begin{equation}
S_2=-\frac{T_{2,0}}{w}+\frac{b_7}{b_3}T_{0,0}\left(\frac{1+w}{w}\right)\\ 
-\left(\frac{b_5}{b_3}\right)^2 T_{0,0}\left(\frac{\ln\abs{w}+(1+w)}{w}\right) 
\end{equation}
\end{subequations}
where $w=-1+2b_3\xi$. (Eq. \eqref{eq5} corrects errors in ref. \cite{b3}.) For the solutions of eq. \eqref{eq4} for $S_n(n = 3 \ldots 6)$ see the appendix.

An alternate expression for the effective action that follows from the conformal anomaly is \cite{b2}
\begin{equation}\label{eq6}
L=-\frac{1}{4}\frac{\lambda_0^2}{\bar{\lambda} ^2(t)}\Phi
\end{equation}
where the running coupling $\bar{\lambda}(t)$ satisfies 
\begin{equation}\label{eq7}
\frac{d\bar{\lambda}(t)}{dt}=\beta(\bar{\lambda(t)})\qquad (\bar{\lambda}(t=0)=\lambda_0)
\end{equation}
Eq. \eqref{eq6} satisfies \eqref{eq1} provided $\mu=\mu_0$ is fixed. In ref. \cite{b3} it is shown that eqs. \eqref{eq3} and \eqref{eq6} are consistent provided
\begin{equation}\label{eq8}
T_{n,0}=-\frac{1}{4}\delta_{n,0}.
\end{equation}
Furthermore, these two equations show that
\begin{equation}\label{eq9}
\bar{\lambda}^2(t)=\frac{-\lambda_0^2}{4}\left[\sum_{n=0}^{\infty}S_n(\lambda_0^2t)\lambda_0^{2n}\right]^{-1}.
\end{equation}
More explicitly, from eqs. (\eqref{eq5},\eqref{eq8},\eqref{eq9}) it follows that 
\begin{align}\label{eq10}
 \bar{\lambda}^2(t) &=\lambda_0^2\bigg[(1-2b_3\lambda_0^2 t)+\lambda_0^2\left(\frac{b_5}{b_3}\ln\abs{-1+2b_3\lambda_0^2t}\right)\\ \nonumber
&+\lambda_0^4\left(\frac{b_7}{b_3} \frac{2b_3\lambda_0^2t}{-1+2b_3\lambda_0^2t}-\left(\frac{b_5}{b_3}\right)^2 \frac{\ln\abs{-1+2b_3\lambda_0^2t}+2b_3\lambda_0^2t}{-1+2b_3\lambda_0^2t} \right)+\ldots\bigg]^{-1}
\end{align}

This rather unusual expression for $\bar{\lambda}^2(t)$ can be composed with what can be obtained directly from eq. \eqref{eq7}. For a lowest order solution, from 
\begin{subequations}
\begin{equation}\label{eq11a}
\frac{d\bar{\lambda}^2(t)}{dt}=b_3\bar{\lambda}^3(t)
\end{equation}
we easily find that
\begin{equation}\label{eq11b}
\bar{\lambda}^2(t)=\frac{\lambda_0^2}{1-2b_3\lambda_0^2t}
\end{equation}
\end{subequations}
while if we go the next order 
\begin{subequations}
\begin{equation}\label{eq12a}
\frac{d\bar{\lambda}(t)}{dt}=b_3\bar{\lambda}^3(t)+b_5\bar{\lambda}^5(t)
\end{equation}
it follows that 
\begin{equation}\label{eq12b}
\frac{d\bar{\lambda}^2}{(\lambda^2_0)^2\left(b_3+b_5\lambda^2_0\right)}=2dt \nonumber
\end{equation}
\end{subequations}
which, when integrated, yields $\bar{\lambda}(t)$ in terms of a Lambert $W-$ function \cite{b6}. 
Eq. \eqref{eq11b} is identical to the lowest order contribution to eq. \eqref{eq10}, while eq. \eqref{eq10}  yields no closed 
form expression when $b_3,b_5$ are non-zero.

However, eq. \eqref{eq10} can be related to what is obtained from a perturbative solution to eq  \eqref{eq7} which is found in the following systematic way. We begin by letting $x=\bar{\lambda}^2$ and
$2b_{2\rho+3}=\beta_{\rho} (\rho=0,1,2 \ldots)$ so that eq. \eqref{eq7} becomes \cite{b7}
\begin{equation}\label{eq13}
\frac{dx}{dt}=x^2(\beta_0+\beta_1x+\beta_2x^2+\ldots)
\end{equation}
If we now rescale $t\to t/\epsilon,x \to\epsilon x$, then make the expansion $x=x_0+\epsilon x_1+\epsilon^2x_2+\ldots$ ($x_n(t=0)=x\delta_{n,0}$)
we find that at successive orders in $\epsilon$,
\begin{subequations}
\begin{equation}\label{eq14a}
\frac{dx_0}{dt}=\beta_0x_0^2
\end{equation}
\begin{equation}\label{eq14b}
\frac{dx_1}{dt}=\beta_0x_0^2+2\beta_1x_0x_1
\end{equation}
\begin{equation}\label{eq14c}
\frac{dx_2}{dt}=\beta_0(x_1^2+2x_0x_2)+3\beta_1x_1x_0^2+\beta_4x_0^4  
\end{equation}
\end{subequations} 
Solving these equations in turn leads to 
\begin{subequations}
\begin{equation}\label{eq15a}
x_0=\frac{x}{1-\beta_0xt}
\end{equation}
\begin{equation}\label{eq15b}
x_1=-x^2\frac{\beta_1}{\beta_0}\frac{\ln\abs{1-\beta_0xt}}{(1-\beta_0xt)^2}
\end{equation}
\end{subequations}
etc.\\
The solutions for $x_n (n=2 \ldots 5)$ are given in the appendix.

An alternate approach is to systematically solving  eq. \eqref{eq7} is to write (in analogy with eq. \eqref{eq3} \cite{b8})
\begin{subequations}
\begin{equation}\label{eq16a}
x(\mu_0)=x(\mu)\sum_{n=0}^{\infty}\sum_{m=0}^{\infty}\tau_{n,m}x^n(\mu)\ln^m\left(\mu^2/\mu^2_0\right)
\end{equation}
\begin{equation}\label{eq16b}
\equiv\sum_{n=0}^{\infty}\sigma_n(\zeta)x^{n+1}(\mu)\qquad (\sigma_n(0)=\delta_{n0})
\end{equation}
\end{subequations}
where $\zeta=x(\mu)\ln\left(\mu^2/\mu_0^2\right).$ If now $\beta(x)=x^2\sum_{n=0}^{\infty}\beta_nx^n$ and
\begin{subequations}
\begin{equation}\label{eq17a}
\mu^2\frac{d}{d\mu^2}x(\mu_0)=0
\end{equation}
\begin{equation}\label{eq17b}
\mu^2\frac{d}{d\mu^2}x(\mu)=\beta\left(x(\mu)\right)
\end{equation}
\end{subequations}
then we see that
\begin{subequations}
 \begin{equation}\label{eq18a}
(1+\beta_0\zeta)\sigma_0^{\prime}=-\beta_0\sigma_0
\end{equation}
\begin{equation}\label{eq18b}
(1+\beta_0\zeta)\sigma_1^\prime+2\beta_0\sigma_1=
(-\beta_1\sigma_0-\beta_1\zeta\sigma_0^{\prime})
\end{equation}
\begin{equation}\label{eq18c}
(1+\beta_0\zeta)\sigma_2^\prime+3\beta_0\sigma_2=
(-\beta_2\sigma_0-\beta_2\zeta\sigma_0^{\prime})+(-2\beta_1\sigma_1-\beta_1\zeta\sigma_1^{\prime})
\end{equation}
\end{subequations}
These equations have the solutions
\begin{subequations}
\begin{equation}\label{eq19a}
\sigma_0=(1+\beta_0\zeta)^{-1}
\end{equation}
\begin{equation}\label{eq19b}
\sigma_1=-\left(\frac{\beta_1}{\beta_0}\right)^2\frac{\ln\abs{1+\beta_0\zeta}}{(1+\beta_0\zeta)^2}
\end{equation}
\begin{align}\label{eq19c}
\sigma_2=&\left(\left(\frac{\beta_1}{\beta_0}\right)^2-\frac{\beta_2}{\beta_0}\right)\left(\frac{1}{(1+\beta_0\zeta)^2}-\frac{1}{(1+\beta_0\zeta)^3}\right) \\ \nonumber
&-\left(\frac{\beta_1}{\beta_0}\right)^2\frac{1}{(1+\beta_0\zeta)^3}\left(\ln\abs{1+\beta_0\zeta}-\ln^2\abs{1+\beta_0\zeta}\right)
\end{align}
\end{subequations}
etc.\\
These solutions to eq. (18) are seen to be equivalent to those of eq. (14).

With the solution to eq. \eqref{eq7} given by eq. (15) (or alternatively eq. (19)),
we find that this is equivalent to the expression for the running coupling given by eq. \eqref{eq9} where the running coupling
appearing in eq. \eqref{eq9} is expanded in powers of $\lambda_0^2$. This holds true to the order that we have 
computed ($\lambda_0^{12}$) and we anticipate that it would be true to all orders in 
$\lambda_0^2$.  Eq. \eqref{eq9} is unusual in that the dependence of $\bar{\lambda}^2(t)$ on $t$ is exclusively in the denominator.

The sums $\sum_{n=0}^{\infty} S_n(\lambda^2t) \lambda^{2n}\Phi$ and $\sum_{n=0}^{\infty}\sigma_{n}(\zeta)x^{n+1}$ in eqs. \eqref{eq3} and \eqref{eq16a},\eqref{eq16b} represent leading-log (LL) contributions  (for $n=0$),
next-to-leading-log (NLL) contributions (for $n=1$) and, in general, $N^pLL$ contribution (for $n=p$) for $L$ and $\bar{\lambda}^2$ respectively. It proves possible to use the renormalization group equation to
 perform parts of these sums,
as was done in ref. \cite{Mnogo} when considering the effective potential.

We illustrate this by first considering $\sigma_n(\zeta)$.
From eqs. \eqref{eq16b} and \eqref{eq17a},\eqref{eq17b} we find that 
\begin{equation}\label{eq20}
\left[(1+\beta_0\zeta)\frac{d}{d\zeta}+(n+1)\beta_0\right]\sigma_n+\sum_{\rho=1}^n\beta_{\rho}\left[\zeta\frac{d}{d\zeta}+(n+1-\rho)\right]\sigma_{n-\rho}=0
\end{equation}
(This generalizes eq. \eqref{eq18a},\eqref{eq18b},\eqref{eq18c}.) The general form of $\sigma_n(\zeta)$ that follows from eq. \eqref{eq20} is 
\begin{equation}\label{eq21}
\sigma_n=\sum_{i=0}^n\sum_{j=0}^i\sigma_{i,j}^n\frac{L^j}{U^{i+1}}
\end{equation}
where $U=1+\beta_0\zeta$ and $L=\ln U$. Substitution of eq. \eqref{eq21} into eq. \eqref{eq20} leads to the recursion relation
\begin{align}\label{eq22}
\beta_0\left[(j+1) \sigma_{i,j+1}^n+(n+1-i)\sigma_{i,j}^n\right]
+\sum_{\rho=1}^n  \beta_{\rho}\bigg[-(j+1)\sigma_{i-1,j+1}^{n-\rho}+(i-1)
\sigma_{i-1,j}^{n-\rho}\\ \nonumber
+(j+1)\sigma_{i,j+1}^{n-\rho}-i\sigma_{i,j}^{n-\rho}+(n+1-\rho)\sigma_{i,j}^{n-\rho}\bigg] = 0.
\end{align}
If in eq. \eqref{eq22} we set $i=n+1$, then
\begin{equation}\label{eq23}
\sigma_{n+1,j+1}^n=\rho_1\left[\frac{n}{j+1}\sigma_{n,j}^{n-1}-\sigma_{n,j+1}^{n-1}\right]
\end{equation}
where $\rho_n=-\beta_n/\beta_0.$ If in eq. \eqref{eq23}, we set $j=n-1$, then 
\begin{equation}\label{eq24}
\sigma_{n+1,n}=\rho_1\sigma_{n,n-1}^{n-1}=(\rho_1)^n\sigma_{10}^0=(\rho_1)^n
\end{equation}
as by eq. \eqref{eq19a}, $\sigma_{10}^0=1.$
Restricting $\sigma_{ij}^n$ in eq. \eqref{eq21} to $\sigma_{n,n+1}^n$, we find from eq. \eqref{eq16b} that
\begin{align}\label{eq25}
x(\mu_0) &=\sum_{n=0}^{\infty}\rho_1^n\frac{L^n}{U^{n+1}}x^n(\mu) \\ \nonumber
&=\frac{x(\mu)}{U-\rho_1Lx(\mu)}
\end{align}
or, more explicitly (reversing the roles of $\mu$ and $\mu_0$)
\begin{equation}\label{eq26}
x(\mu)=\frac{x(\mu_0)}{1-\beta_0\ln\left(\frac{\mu^2}{\mu_0^2}\right)+\frac{\beta_1}{\beta_0}\ln\left(1-\beta_0\ln\left(\frac{\mu^2}{\mu_0^2}\right)\right)x(\mu_0)}
\end{equation}
which is consistent with eq. \eqref{eq10}.

If $j=n-2$ in eq. \eqref{eq23}, an explicit expression for $\sigma_{n+1,n-1}^n$ can be found following the approach of ref. \cite{b5};
this further modifies the expression for $x(\mu)$ in eq. \eqref{eq26}.

In a similar fashion, one can use eq. \eqref{eq4} to see that 
\begin{equation}\label{eq27}
S_n(\xi)=\sum_{i=0}^n\sum_{j=0}^i S_{ij}^n\frac{L^j}{w^{i-1}};
\end{equation}
in analogy with eq. \eqref{eq22} we find that 
\begin{align}\label{eq28}
(j+1)S_{i,j+1}^n+(n-i)S_{ij}^n+\sum_{\rho=1}^{n-1}\chi_{2\rho+3}\bigg[(j+1)S_{i-1,j+1}^{n-\rho}-(i-2)S^{n-\rho}_{i-1,j}\\ \nonumber
+(j+1)S^{n-\rho}_{i,j+1}+(n-\rho-i)S_{ij}^{n-\rho}\bigg]=0,
\end{align}
where $\chi_{2\rho+3}=b_{2\rho+3}/b_3 \;\,(\rho=1,2\ldots).$
For $i=n$ and $j=n-1$, eq. \eqref{eq28} reduces to 
\begin{equation}\label{eq29}
S_{n,n}^n-\chi_5\frac{(n-2)}{n}S_{n-1,n-1}^{n-1}=0.
\end{equation}
As $S_{0,0}^0=\frac{1}{4}$ (by eqs. \eqref{eq5a},\eqref{eq8}), we see by eq. \eqref{eq29} that $S_{1,1}^1=-\chi_5/4,S_{n,n}^n=0\;(n\ge2).$ If we only consider the contributions to $S_n$ coming from $S_{n,n}^n,$
 it follows from eq. \eqref{eq9} that
\begin{equation}\label{eq30}
\bar{\lambda}^2(t)=-\frac{\lambda_0^2}{4}\left[\frac{1}{4}w-\frac{\chi_5}{4}(\ln w)\lambda_0^2\right]^{-1}
\end{equation}
which is identical to eq. \eqref{eq26}. 

Further results that follow from eq. \eqref{eq28} are
\begin{subequations}
\begin{equation}\label{eq31a} 
S^2_{2,0}=-\frac{1}{4} (\chi_7 - \chi_5^2), 
\end{equation}
\begin{equation}\label{eq31b} 
S_{3,1}^3= -\frac{\chi_5\chi_7}{4}
\end{equation}
\begin{equation}\label{eq31c}
S^n_{n,n-2}=-\frac{\chi_5^{n-2}\chi_7}{4}-\frac{\chi_5^n}{4}\left(\frac{1}{2}+\frac{1}{3}+\ldots+\frac{1}{n-2}\right)\;\;(n \geq 4)
\end{equation}
\end{subequations}
\begin{equation}\label{eq32}
S^n_{n-1,n-1}=0 \;\; (n \geq 1)
\end{equation}
\begin{subequations}
\begin{equation}\label{eq33a}
S^2_{1,0} = \frac{\chi_5^2}{4}-\frac{\chi_7}{4},
\end{equation}
\begin{equation}\label{eq33b}
S^3_{2,1} =0
\end{equation}
\end{subequations}
\begin{equation}\label{eq34}
S^n_{n-1,n-3}= \frac{1}{4}  \left( \chi_7 \chi_5^{n-2} - \chi_5^{\,n}\right)\; (n\geq 3).
\end{equation}
These contributions to $L$ in eq. \eqref{eq3} can now be easily summed. (For the contribution of eq. \eqref{eq31c} see the appendix.)
The final result for $L/\Phi$ coming from eqs. (31-34) is the following
\begin{align}
 L/\Phi=\frac{1}{4}\bigg[w-\chi_5\ln w\lambda^2+(\chi_5^2-\chi_7)\left(\frac{1+w}{w}\right)\lambda^4 \\ \nonumber
-\frac{\lambda^4}{w}\frac{1}{1-\lambda^2\ln w/w}\left(\lambda^2\ln w/w-\ln\left(1-\lambda^2\ln w/w\right)\right)\\ \nonumber
-\frac{\lambda^6}{w}\left(\chi_7\chi_5-\chi_5^3\right)\left(1+\lambda^2\chi_5\ln w/w\right)^{-1}
\bigg]
\end{align}

where the last two terms receive contributions from all terms of order $N^pLL$. 
\section{Discussion}
By exploiting the conformal anomaly, the effective action for a constant external gauge field can be expressed in terms of the running coupling. We have used this result to find an alternative expression 
for the running coupling that is perturbatively equivalent to the usual solutions to eq. \eqref{eq7}.

We have also shown how portions of all $N^pLL$ contributions to the running coupling can be summed.

In \cite{Knie} a different approach was used to integrate eq. (7). In this reference, one takes
\begin{equation}\label{eq36}
t = \int \frac{d\lambda}{b_3\lambda^3 + b_5\lambda^5 + \ldots} = 
-\frac{1}{2b_3} \left[\frac{1}{\lambda^2} + \frac{b_5}{b_3}\ell n \lambda^2 + \left(\frac{b_7}{b_3} - \frac{b_5^2}{b_7^2}\right)\lambda^2 + \ldots 
\right]
\end{equation}
which is obtained by expanding the denominator of the integral.  This is now solved interatively to yield
\begin{equation}\label{eq37}
\lambda^2 = -\frac{1}{2b_3t} + \frac{b_5}{4b_3^3}\frac{1}{t^2} \ell n \left(-\frac{1}{2b_3t}\right) + \ldots 
\end{equation}
A systematic to using \eqref{eq36} to expand $\lambda^2$ in powers of $t^{-1}$ and $\ell n\,t$ is given in \cite{Peng}; the techniques used resemble those that lead to \eqref{eq35} above.  However, the renormalization group equation is not employed directly in ref. [11] as it is here.

%\vspace{.3cm}.\\

\bigskip

\noindent
{\Large\bf{Acknowledgment}}\vspace{.5cm}\\
Roger Macleod had a helpful suggestion.

\newpage

\noindent
{\Large\bf{Appendix}}\\

The solutions for $x_n(n = 2 
\ldots 5)$ are as follows:
\begin{subequations}
\begin{align}
x_2=\frac{1}{\beta _0^2 w^3} \bigg[x^3 \left(\beta _1^2 \left(w-\ln^2 w+\ln (w)+1\right)-\beta _0 \beta _2 (w+1)\right)\bigg]
\end{align}
\begin{align}
x_3=-\frac{1}{2 \beta _0^3 w^4} x^4 \bigg[\beta _0^2 \beta _3 \left(w^2-1\right)+&\beta _1^3 \left((w+1)^2+2 \ln^3 w-5 \ln^2 w-4 (w+1) \ln w\right) \\ \nonumber
&-2 \beta _0 \beta _2 \beta _1 (w (w+1)-(2 w+3) \ln  (w))\bigg]
\end{align}
\begin{align}
x_4=\frac{1}{6 \beta _0^4 w^5} x^5 & \bigg[-2 \beta _0^2 \left(\beta _0 \beta _4 \left(w^3+1\right)-\beta _2^2 (w-5) (w+1)^2\right) \\ \nonumber
&-6 \beta _0 \beta _2 \beta _1^2 \left(-\left(2 w^2+5 w+3\right) \ln w+(w-3) (w+1)^2+3
   (w+2) \ln^2 w\right)\\ \nonumber
&+\beta _1^4 \left(-6 \left(w^2+5 w+4\right) \ln w+(w+1)^2 (2 w-7)-6 \ln^4 w+26 \ln^3 w+9 (2 w+1) \ln^2 w\right)\\ \nonumber
&+\beta _0^2 \beta _3 \beta _1 \left(4 w^3+3
   w^2-6 \left(w^2-2\right) \ln w+1\right)\bigg]
\end{align}
\begin{align}
x_5 &=-\frac{1}{12 \beta _0^5 w^6} x^6 \bigg[\beta _1^5 \big(6 \left(3 w^2+26 w+23\right) \ln^2 w+(w+1)^3 (3 w-17)+12 \ln^5 w-77 \ln^4 w \\ \nonumber
+& (22-48 w) \ln^3 w-2 (w+1)^2 (4 w-11) \ln w\big)+3 \beta _0^3 \left(\beta _0 \beta _5 \left(w^4-1\right)-2 \beta _2 \beta _3 \left(-w^2+w+2\right)^2\right) \\ \nonumber
+&\beta _0^2 \beta _3 \beta _1^2 \big(\left(9 w^2-22 w+23\right) (w+1)^2+6 \left(3 w^2-10\right)
   \ln^2 w-2 \left(8 w^3+15 w^2-7\right) \ln w\big) \\ \nonumber
-&6 \beta _0 \beta _2 \beta _1^3 \big((w+1)^2 \left(2 w^2-8 w-3\right)+\left(6 w^2+26 w+27\right) \ln^2 w+\left(-4 w^3+2 w^2+30
   w+24\right) \ln w \\ \nonumber
-&4 (2 w+5) \ln^3 w\big)+\beta _0^2 \beta _1 \big(2 \beta _0 \beta _4 \left(-3 w^4-2 w^3+2 \left(2 w^3+5\right) \ln w+1\right)\\ \nonumber
&+\beta _2^2 (w+1) \left(9 w^3-29w^2+\left(-8 w^2+44 w+100\right) \ln w-37 w+1\right)\big)\bigg]
\end{align}
\end{subequations}

The solutions for $S_n (n = 3 \ldots 6)$ are as follows:
\begin{subequations}
\begin{align}
S_3=-\frac{1}{8 w^2}\bigg[ \chi _9 \left(w^2-1\right)-2 \chi _7 \chi _5 \left(w^2+w-\ln (w)\right)+\chi _5^3 \left((w+1)^2-\ln^2 w\right)\bigg]
\end{align}
\begin{align}
S_4&=\frac{1}{24 w^3}\bigg[ -2 \chi _{11} \left(w^3+1\right)+\chi _9 \chi _5 \left(4 w^3+3 w^2+6 \ln (w)+1\right)+2 \chi _7^2 (w-2) (w+1)^2 \\ \nonumber
&-6 \chi _7 \chi _5^2 \left((w-1) (w+1)^2+\ln^2 w-(w+1) \ln
   (w)\right) \\ \nonumber
&+\chi _5^4 \left((w+1)^2 (2 w-1)+2 \ln^3 w-3 \ln^2 w-6 (w+1) \ln (w)\right)\bigg]
\end{align}
\begin{align}
S_5&= -\frac{1}{48 w^4}\bigg[ \chi _{13} \big(-6 \chi _7 \chi _5^3 \big((w+1)^2 \left(2 w^2-2 w-1\right)+\left(-2 w^2+2 w+4\right) \ln (w)-2 \ln^3 w\\ \nonumber
&+(2 w+5) \ln^2 w\big)-3 \left(2 \chi _7 \chi _9
   \left(w^4-w^2+2 w+2\right)-\chi _{13} \left(w^4-1\right)\right) \\ \nonumber
&+\chi _9 \chi _5^2 \left(9 w^4+8 w^3-6 \left(w^2-1\right) \ln (w)+12 w-18 \ln^2 w+11\right) \\ \nonumber
&+\chi _5 \left(2 \chi
   _{11} \left(-3 w^4-2 w^3+6 \ln (w)+1\right)+\chi _7^2 (w+1) \left(9 w^3-5 w^2-13 w+24 \ln (w)+1\right)\right)\\ \nonumber
&+\chi _5^5 \left((w+1)^3 (3 w-5)-3 \ln^4 w+10 \ln^3 w+12 (w+1) \ln
   ^2(w)-6 (w+1)^2 \ln (w)\right)\big)\bigg]
\end{align}
\begin{align}
S_6&=\frac{1}{240 w^5}\bigg[ -10 \chi _7 \chi _5^4 \big((w+1)^3 \left(6 w^2-12 w+7\right)+\left(6 w^2-3 w-9\right) \ln^2 w+6 \ln^4 w\\ \nonumber
&-2 (3 w+13) \ln^3 w-6 (w-4) (w+1)^2 \ln (w)\big)+\chi _5^6 \big(3
   (w+1)^3 \left(4 w^2-7 w-1\right)\\ \nonumber
&+30 \left(w^2+5 w+4\right) \ln^2 w+12 \ln^5 w-65 \ln^4 w-30 (2 w+1) \ln^3 w\\ \nonumber
&-10 (w+1)^2 (2 w-7) \ln (w)\big)+\chi _9 \chi _5^3 \big(30
   \left(w^2-5\right) \ln^2 w+3 (w+1)^2 \left(16 w^3-17 w^2+8 w+1\right)\\ \nonumber
&-10 \left(4 w^3+3 w^2+18 w+19\right) \ln (w)+120 \ln^3 w\big) \\ \nonumber
&-2 \big(6 \chi _{15} \left(w^5+1\right)+2 \chi _7^3
   (w+1)^3 \left(3 w^2-9 w+13\right)+2 \chi _{11} \chi _7 \left(-6 w^5+5 w^3+15 w+14\right)\\ \nonumber
&-3 \chi _9^2 \left(2 w^5+5 w^2-3\right)\big)+\chi _5^2 \bigg(\chi _7^2 \big(3 \left(24 w^3-33 w^2+2 w+39\right) (w+1)^2 \\ \nonumber
&-20 \left(w^3-9 w^2-15 w-5\right) \ln (w)-60 (3 w+4) \ln^2 w\big)+2 \chi _{11} \big(10 \left(w^3+1\right) \ln (w)\\ \nonumber
&+3 \left(-6 w^5-5 w^4+10 w+9\right)-60 \ln^2(w)\big)\bigg)+\chi _5 \big(3 \chi _{13} \left(8 w^5+5 w^4+20 \ln (w)+3\right)\\ \nonumber
&-2 \chi _7 \chi _9 (w+1) \left(36 w^4-21 w^3-14 w^2+29 w+30 (w-4) \ln (w)-14\right)\big)\bigg]
\end{align}
\end{subequations}

We also employ, in evaluating the contributions of eq. \eqref{eq31c} to $L$, the result 
\begin{subequations}
\begin{align}
\sum_{n=4}^\infty x^n \left(\frac{1}{2} + \frac{1}{3} + \frac{1}{4} + \ldots + 
\frac{1}{n-2}\right) &= \frac{1}{2} (x^4 + x^5 + x^6 + \ldots) + \frac{1}{3} (x^5 + x^6 + \ldots)\nonumber \\
&= \frac{1}{2} \frac{x^4}{1-x}
 + \frac{1}{3} \frac{x^5}{1-x} + \ldots \nonumber \\
&= -\frac{x^2}{1-x} (x + \ln(1-x)).
\end{align}
\end{subequations}

\end{document}